\documentclass[
 twocolumn, 
 reprint,
 superscriptaddress,
 amsmath,
 amssymb,
 aps,
 pra, 
]{revtex4-2}

\usepackage{graphicx}
\usepackage{dcolumn}
\usepackage{multirow} 
\usepackage{xcolor}
\usepackage[normalem]{ulem}
\usepackage{bm}
\usepackage{hyperref}
\usepackage{physics}
\usepackage{braket}
\usepackage{amsmath}
\usepackage{amsfonts}
\usepackage{amssymb}
\usepackage{amsthm}
\usepackage{mathtools}
\usepackage{color} 
\usepackage{gensymb}
\usepackage{here}

\begin{document}

\title{Impact of molecular orbital localization on quantum computational resources for Hamiltonian simulation: A benchmark study of hydrogen chain systems}

\author{Kenji Sugisaki}
\email{kensugisaki@tohmatsu.co.jp}
\affiliation{Deloitte Tohmatsu LLC, 3-2-3 Marunouchi, Chiyoda-ku, Tokyo 100-8363, Japan}

\author{Yuhei Tachi}
\affiliation{Deloitte Tohmatsu LLC, 3-2-3 Marunouchi, Chiyoda-ku, Tokyo 100-8363, Japan}

\author{Masayoshi Terabe}
\affiliation{Deloitte Tohmatsu LLC, 3-2-3 Marunouchi, Chiyoda-ku, Tokyo 100-8363, Japan}

\author{Hiroyuki Tezuka}
\affiliation{Deloitte Tohmatsu LLC, 3-2-3 Marunouchi, Chiyoda-ku, Tokyo 100-8363, Japan}

\date{\today}
    
\begin{abstract}
We investigate how molecular orbitals used as the basis of wave function expansion and how operator coefficient-based and locality-based Hamiltonian truncation affects the computational cost of Trotter decomposition-based Hamiltonian simulation in one-dimensional hydrogen chain systems. The analysis is performed using both Hartree--Fock canonical molecular orbitals (CMOs) and Pipek--Mezey-based localized molecular orbitals (LMOs). For short hydrogen chains, we evaluate the ground-state energy and fidelity and find that, in the CMO-based wave function expansion, introducing a threshold on Hamiltonian coefficients is effective in reducing the gate cost while maintaining computational accuracy. In contrast, in the LMO-based wave function expansion, operator locality-based Hamiltonian truncation is found to be more effective. By fitting the relationship between the truncation threshold and the ground-state energies and fidelities with empirical formulas, we estimate the threshold values required to achieve high fidelity ($F \ge 0.99$) in the ground-state wave function. Using the estimated thresholds, we then perform quantum gate resource estimation for longer hydrogen chains up to H$_{100}$. The results suggest an exponential advantage of the LMO-based wave function expansion with Hamiltonian truncation: the number of quantum gates required for Hamiltonian simulation grows polynomially when the CMO-based wave function expansion with operator coefficient-based Hamiltonian truncation is adopted, whereas it grows polylogarithmically when the LMO-based wave function expansion is combined with operator locality-based Hamiltonian truncation. These results provide useful guidelines for choosing orbital representations and Hamiltonian truncation strategies in large-scale quantum chemical simulations.
\end{abstract}

\maketitle

\section{Introduction}

Quantum computers have attracted considerable attention as a next-generation computational platform and are expected to have broad applications in fields such as chemistry, combinatorial optimization, machine learning, and communication technologies. Among these potential applications, quantum chemical calculations for determining the electronic structure of atoms and molecules are regarded as one of the most promising targets, owing to their importance in materials discovery, catalysis, and drug design~\cite{Cao-2019, Bauer-2020, McArdle-2020, Motta-2022, Blunt-2022}. To address such problems, a variety of quantum algorithms have been proposed, including quantum phase estimation (QPE)~\cite{Kitaev-1995, Abrams-1999, Aspuru-Guzik-2005}, variational quantum eigensolver (VQE)~\cite{Peruzzo-2014, Tilly-2022}, and quantum-selected configuration interaction (QSCI)~\cite{Kanno-2023, Robledo-Moreno-2025, Sugisaki-2025}.

In anticipation of the era of fault-tolerant quantum computing (FTQC), estimating the computational resources required to perform practically relevant quantum chemical simulations has become an important research objective. Resource estimation provides essential guidance for assessing the feasibility of large-scale applications and for identifying bottlenecks in algorithmic design. Previous studies have primarily investigated the dependence of quantum computational cost on the choice of simulation method, such as Trotterization or qubitization \cite{Babbush-2018, Ku-2025}, second quantization or first quantization \cite{Su-2021}, as well as on specific Hamiltonian-factorization or compression strategies, including quantum complex exponential least squares (QCELS) \cite{Nelson-2024}, double factorization tensor hypercontraction (DFTHC) \cite{vonBurg-2021, Lee-2021, Low-2025}, sum-of-squares spectral amplification (SOSSA) \cite{Low-2025, King-2026}, block-invariant symmetry shift (BLISS) \cite{Loaiza-2023, Caesura-2025}, etc \cite{Rubin-2023, Rowe-2026}. Resource requirements have also been examined for chemically important target systems, such as FeMo cofactor (FeMoco) in the active site of nitrogenase \cite{Reiher-2017, Li-2019, Gunther-2026} and iron--sulfur complexes [Fe$_n$S$_n$] as model systems of FeMoco \cite{Kanasugi-2026}, cytochrome P450 \cite{Goings-2022}, and so on \cite{Otten-2024, Bellonzi-2024, Gu-2025}, in order to clarify the molecular dependence of quantum simulation costs. In addition to such investigations into the choice of simulation method and target molecule, efforts have also been made to develop software tools that support resource estimation and fault-tolerant compilation, including TFermion \cite{Casares-2022}, QREChem \cite{Otten-2023}, the Azure Quantum Resource Estimator \cite{Beverland-2022}, Qualtran \cite{Harrigan-2024}, BenchQ \cite{BenchQ}, and FTCircuitBench \cite{Harkness-2026}

By contrast, relatively little attention has been paid to how the choice of molecular orbitals used for wave function expansion affects quantum computational resources. In conventional quantum chemistry, Hartree–Fock (HF) canonical molecular orbitals (hereafter denoted as CMOs) or natural orbitals obtained from post-HF calculations such as the second-order M{\o}ller--Plesset (MP2) \cite{Jensen-1988} are often employed as the one-electron basis for correlated calculations. However, CMOs and natural orbitals are often delocalized over the entire molecule, which leads to broadly distributed two-electron interactions and, in the second-quantized Hamiltonian, effectively all-to-all couplings among orbitals. 

In traditional quantum chemistry, especially for large molecular systems like proteins, localized molecular orbitals (denoted as LMOs) are widely used \cite{Neese-2009, Riplinger-2016, Amor-2021}. A key advantage of orbital localization is that interactions between spatially well-separated orbitals often become negligibly small and can therefore be truncated with limited loss of accuracy. This locality has long been exploited to reduce computational cost in electronic structure methods. From the viewpoint of quantum computation, orbital locality may likewise provide a route to reducing Hamiltonian complexity and, consequently, the resources required for simulation. In addition, although the wave function of a strongly correlated system exhibits multiconfigurational character in the Slater determinant basis, it can in some cases be well approximated by a single configuration state function (CSF) in the LMO basis \cite{Sugisaki-2016, Izsak-2023}. It should be also noted that quantum circuits for efficiently prepare specific CSF on a quantum computer have already been reported \cite{Sugisaki-2016, Sugisaki-2019, Marti-Dafcik-2025}. 

Indeed, several studies in the quantum computing literature have suggested that the choice of orbital basis can have important consequences for algorithmic performance \cite{Babbush-2015, Koridon-2021, Izsak-2023b, Besserve-2024, Materia-2024, Sugisaki-2024a, Sugisaki-2024b, Tachi-2025, Shajan-2025, Merz-2026}. In particular, for quantum chemical computations based on Trotter decomposition such as VQE with unitary coupled cluster ansatz \cite{Anand-2022} and QPE, Trotter errors can lead to a breakdown of size consistency, and it has been reported that the use of LMOs can mitigate this violation \cite{Sugisaki-2024a, Sugisaki-2024b}. These observations indicate that orbital localization is not merely a technical detail of basis representation, but a potentially important factor in determining the accuracy and efficiency of quantum simulations.

In this work, we investigate the impact of molecular orbital localization on quantum computational resources for Trotter decomposition-based Hamiltonian simulation, by focusing on linear hydrogen chain systems as a benchmark platform. Hydrogen chains provide a conceptually simple yet nontrivial model for examining the interplay between orbital representation, Hamiltonian sparsity, and wave function accuracy. Specifically, we compare CMO- and LMO-based wave function expansions and examine how Hamiltonian truncation method and threshold influences the number of retained Hamiltonian terms, ground-state energy, and the fidelity of the ground-state wave function. Through this analysis, we aim to clarify how orbital localization affects the trade-off between Hamiltonian compression and the preservation of physically relevant quantum states, thereby providing insight into basis-design strategies for future fault-tolerant quantum chemical simulations.

\section{Computational conditions}

\begin{figure*}[t]
    \centering   \includegraphics[width=\textwidth]{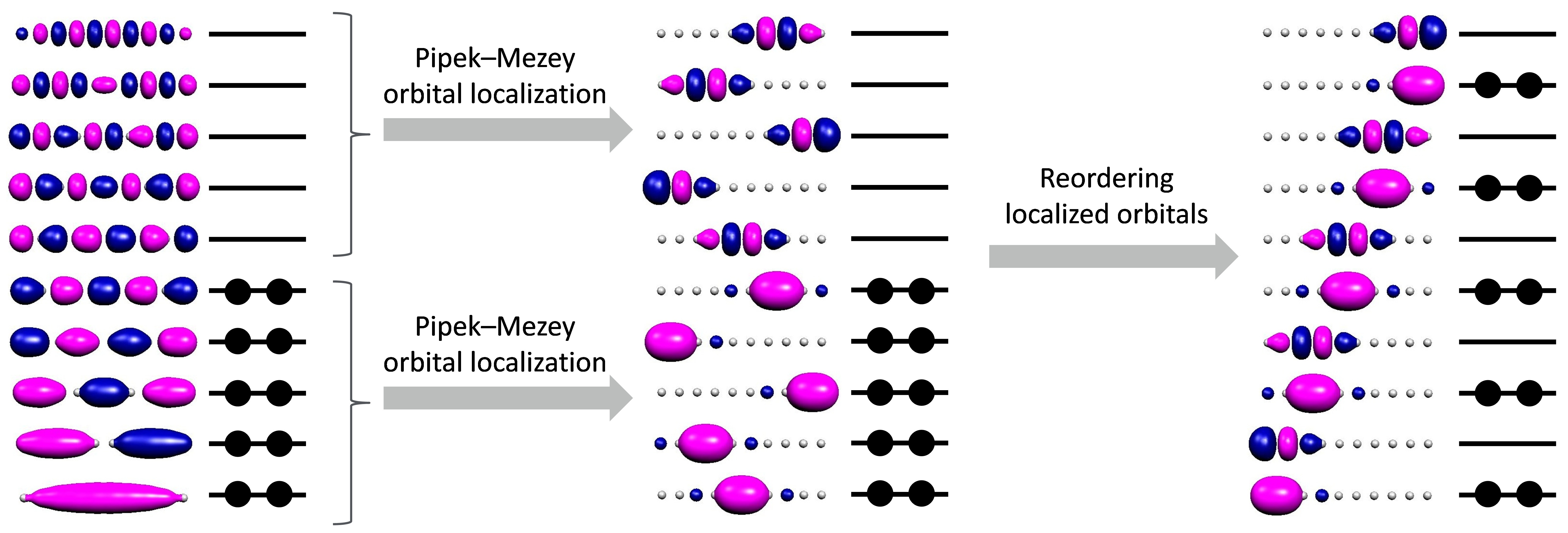}
    \caption{Procedure of orbital localization, in the case of H$_{10}$. Black circles represent the electron occupancies in the RHF wave function. }
    \label{fig:fig1}
\end{figure*}

In this study, we focus on one-dimensional linear hydrogen chain systems, H$_n$, with interatomic distances of 1.0 \AA, where $n$ = 8, 10, 12, 14, 16, 18, 20, 30, 40, 50, 60, 70, 80, 90, and 100. The molecular orbital localization procedure is summarized in Figure \ref{fig:fig1}. First, the RHF/STO-3G \cite{STO-3G} calculation is performed. After that, LMOs are generated by applying the Pipek--Mezey localization method \cite{Pipek-1989} for occupied and unoccupied orbitals separately, to make the RHF wave function invariant. The resulting LMOs are then reordered according to their relative positions. Note that a similar LMO reordering scheme was employed in Hamiltonian simulation of one-dimensional $\pi$-conjugated carbon chain molecules known as carbynes \cite{Sugisaki-2025}, where operator locality-based Hamiltonian truncation was combined with matrix product operator (MPO)-based classical optimization of quantum circuits for Hamiltonian simulation.  
The RHF and orbital localization calculations are carried out using the PySCF library \cite{PySCF}. The one- and two-electron integrals, $h_{pq}$ and $g_{pqrs}$, appearing in the second-quantized Hamiltonian in eqn (\ref{eq1}), are then computed and stored in FCIDUMP format. 
\begin{eqnarray}
    H = \sum_{pq\sigma}h_{pq} a_{p\sigma}^\dagger a_{q\sigma} + \frac{1}{2}\sum_{pqrs\sigma\sigma'} g_{pqrs} a_{p\sigma}^\dagger a_{q\sigma'}^\dagger a_{s\sigma'} a_{r\sigma}
    \label{eq1}
\end{eqnarray}
Here, $p$, $q$, $r$, and $s$ label spatial orbitals, and $\sigma$ and $\sigma'$ represent the spin degree of freedom ($\sigma \in \{\alpha, \beta\}$). The qubit Hamiltonians $H_\text{qub}$ defined in eqn (\ref{eq2}) are then constructed by applying Jordan--Wigner (JW) transformation \cite{Jordan-1928}, unless otherwise stated, using OpenFermion library \cite{OpenFermion}. 
\begin{eqnarray}
    H_\text{qub} = \sum_j c_j P_j
    \label{eq2}
\end{eqnarray}
Here, $P_j$ is a tensor product of Pauli operators defined in eqn (\ref{eq3}), termed as a Pauli string.
\begin{eqnarray}
    P_j = \bigotimes_u \sigma_u,\ \ \ \sigma_u \in\{I_u, X_u, Y_u, Z_u\} 
    \label{eq3}
\end{eqnarray}

In the JW, each qubit encodes the electron occupation number of the corresponding spin-orbital, i.e., $\ket{1}$ if the spin-orbital is occupied, and $\ket{0}$ otherwise. Note that there are two mainly used spin-orbital ordering conventions, namely interleaved arrangement ($\ket{\psi} = \ket{\phi_{1\alpha} \phi_{1\beta} \phi_{2\alpha} \phi_{2\beta}\cdots}$) and block spin arrangement ($\ket{\psi} = \ket{\phi_{1\alpha} \phi_{2\alpha} \cdots \phi_{1\beta} \phi_{2\beta}\cdots}$). In the present study, we adopt the block spin arrangement because it simplifies the analysis based on operator locality. 

Hamiltonian truncation by discarding unimportant operators has the potential to significantly reduce quantum computational cost. In this work, we consider two approaches to Hamiltonian truncation, namely operator coefficient-based and operator locality-based strategies. In the former approach, we discard terms in the second-quantized Hamiltonian when the absolute values of their coefficients are smaller than a threshold $c_\text{thre}$. 
\begin{eqnarray}
    H_{c_\text{thre}} = \smashoperator{\sum_{|h_{pq}| \ge c_{\text{thre}}}} h_{pq} a_{p\sigma}^\dagger a_{q\sigma} + \frac{1}{2} \smashoperator{\sum_{|g_{pqrs}|\ge c_{\text{thre}}}} g_{pqrs} a_{p\sigma}^\dagger a_{q\sigma'}^\dagger a_{s\sigma'} a_{r\sigma}
    \label{eq4}
\end{eqnarray}

In the latter approach, we discard terms in the qubit Hamiltonian whose operator locality $k$ exceeds a predefined threshold $k_\text{max}$. The operator locality $k$ is defined as the number of Pauli-X, Y, and Z operators appearing in a given Pauli string. However, if the Hamiltonian truncation is performed naively by imposing a threshold on the locality of the qubit Hamiltonian, a bias in the treatment of spin configurations may arise. This is because, for two-electron terms, locality of the qubit operator differs between cases in which two spin-$\alpha$ electrons or two spin-$\beta$ electrons are excited and cases in which one spin-$\alpha$ and one spin-$\beta$ electrons are excited. Therefore, in the present work, for each two-electron term, we define the locality $k$ based on the case in which one spin-$\alpha$ and one spin-$\beta$ electrons are excited, and discard Hamiltonian terms with $k > k_\text{max}$. The locality of the two-electron term $\frac{1}{2}g_{pqrs}a_{p\sigma}^\dagger a_{q\sigma'}^\dagger a_{s\sigma'} a_{r\sigma}$ is calculated using the following equation.
\begin{eqnarray}
    k = |p-r| + |q-s| + 2
    \label{eq5}
\end{eqnarray}

To assess the effect of Hamiltonian truncation, we perform ground-state calculations at the full configuration interaction (full-CI) level for H$_8$, H$_{10}$, H$_{12}$, and H$_{14}$, and at density matrix renormalization group (DMRG) for H$_8$, H$_{10}$, H$_{12}$, H$_{14}$, H$_{16}$, H$_{18}$, and H$_{20}$, using both the original (untruncated) and truncated Hamiltonians. Unless otherwise stated, the bond dimension for DMRG is set to $D$ = 1600. We then evaluate the ground-state energy error $\Delta E = E_{\text{GS;truncated}} - E_{\text{GS;original}}$ and the ground-state fidelity $F = |\langle \Psi_{\text{GS;truncated}}|\Psi_{\text{GS;original}}\rangle|^2$. Full-CI calculations are performed using the PyCI library \cite{PyCI}, and DMRG are carried out using block2 library \cite{block2}. 

\section{Results and discussion}
\subsection{Performance assessment of the Hamiltonian truncation method in the H$_{10}$ system}
\begin{figure}
    \centering   
    \includegraphics[width=\linewidth]{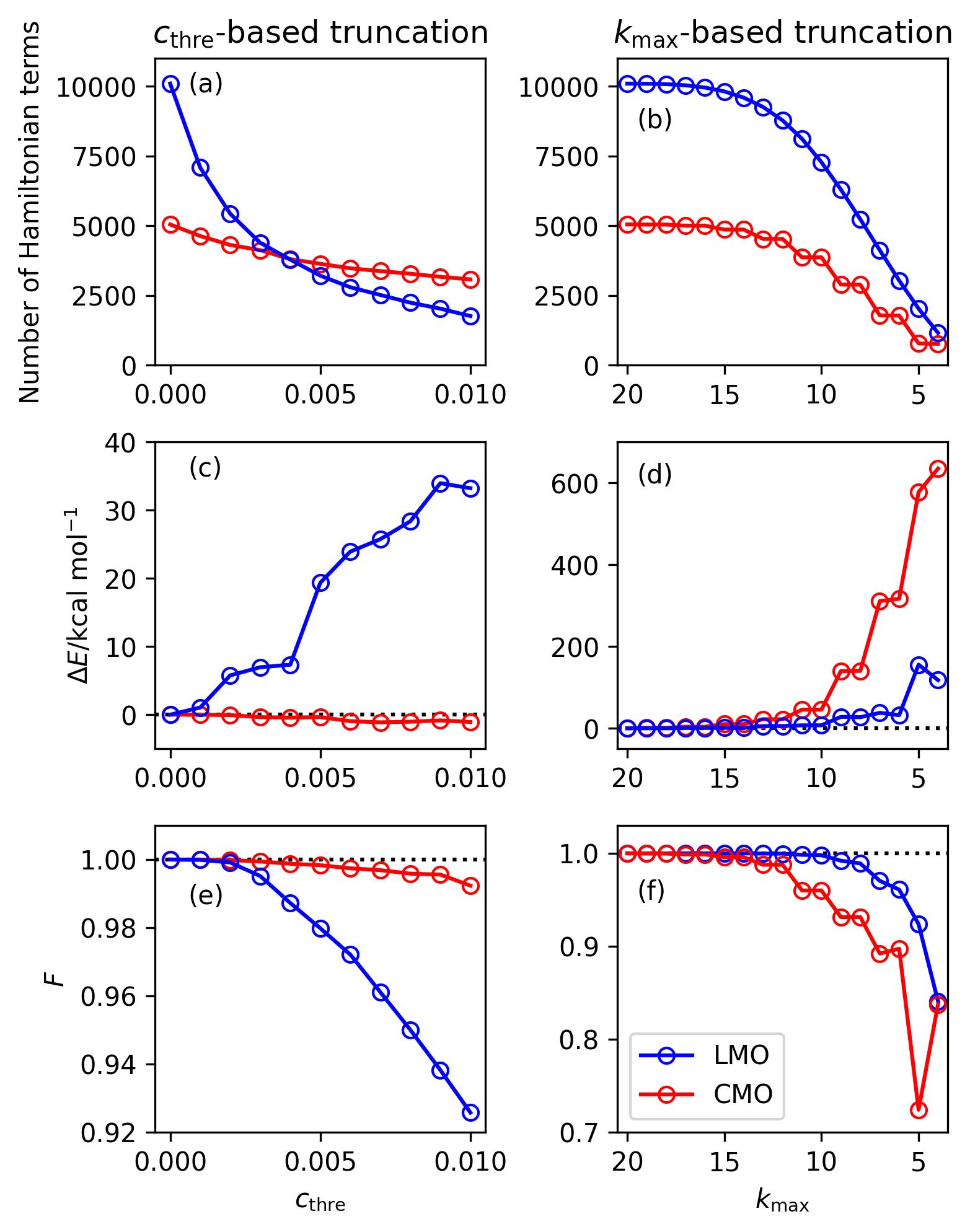}
    \caption{Dependence of Hamiltonian truncation strategies ($c_\text{thre}$-based (a, c, and e) and $k_\text{max}$-based (b, d, and f) approaches) on the number of second-quantized Hamiltonian terms (a and b), ground-state energy error (c and d), and ground-state fidelity (e and f), in the H$_{10}$ system.}
    \label{fig:fig2}
\end{figure}

First, using the H$_{10}$ system as a benchmark, we investigate which of the two Hamiltonian truncation strategies, based on $c_\text{thre}$ or on $k_\text{max}$, is more suitable for the two types of molecular orbitals, CMO and LMO. Figure \ref{fig:fig2} summarizes the dependence of the ground-state energy error $\Delta E$ and ground-state fidelity $F$ at the full-CI level, on the Hamiltonian truncation threshold. 

Figure \ref{fig:fig2} reveals distinct trends in the effectiveness of the Hamiltonian truncation strategies for CMO and LMO. The number of terms in the full Hamiltonian for the LMO basis is approximately twice that for the CMO basis, because orbital localization breaks spatial symmetry. Specifically, the CMO basis belongs to $D_{2h}$ point group in the largest Abelian, whereas the LMO basis belongs to the $C_{2v}$ point group. When $c_{\text{thre}}$-based Hamiltonian truncation is applied in the CMO-basis, $\Delta E$ and $F$ remain close to 0 and 1, respectively, even when a relatively loose threshold value (e.g., $c_\text{thre} = 0.01$) is used. By contrast, the $c_\text{thre}$-based truncation strategy is unsuitable for the LMO basis, as $\Delta E$ and $F$ deviates from 0 and 1, respectively, almost immediately. This also implies that operator coefficient-based randomized compiling strategy such as qDRIFT \cite{Campbell-2019} may not be well suited for Hamiltonian simulation in the LMO basis. 
In contrast, $k_\text{max}$-based truncation is more suitable for the LMO basis than for the CMO basis, as the deviations of $\Delta E$ and $F$ from 0 and 1, respectively, are smaller at low $k_\text{max}$. This trend can be understood from the differences in the spatial distributions of CMOs and LMOs. As shown in Figure \ref{fig:fig1}, CMOs are delocalized over the entire molecule, making the Hamiltonian fundamentally all-to-all in connectivity. Consequently, Hamiltonian terms with large $k$ are not necessarily less important for describing the ground-state wave function. In contrast, LMOs are spatially confined, and electron correlations between distant molecular orbitals are weak. Because we reordered the LMOs according to their relative spatial positions, Hamiltonian truncation based on operator locality approximately corresponds to truncation based on spatial distance. Hereafter, we adopt $c_{\text{thre}}$-based Hamiltonian truncation for the CMO basis, and $k_\text{max}$-based truncation for the LMO basis. 

\subsection{Hydrogen chain length dependence of Hamiltonian term truncation}

\begin{figure}
    \centering   
    \includegraphics[width=0.9\linewidth]{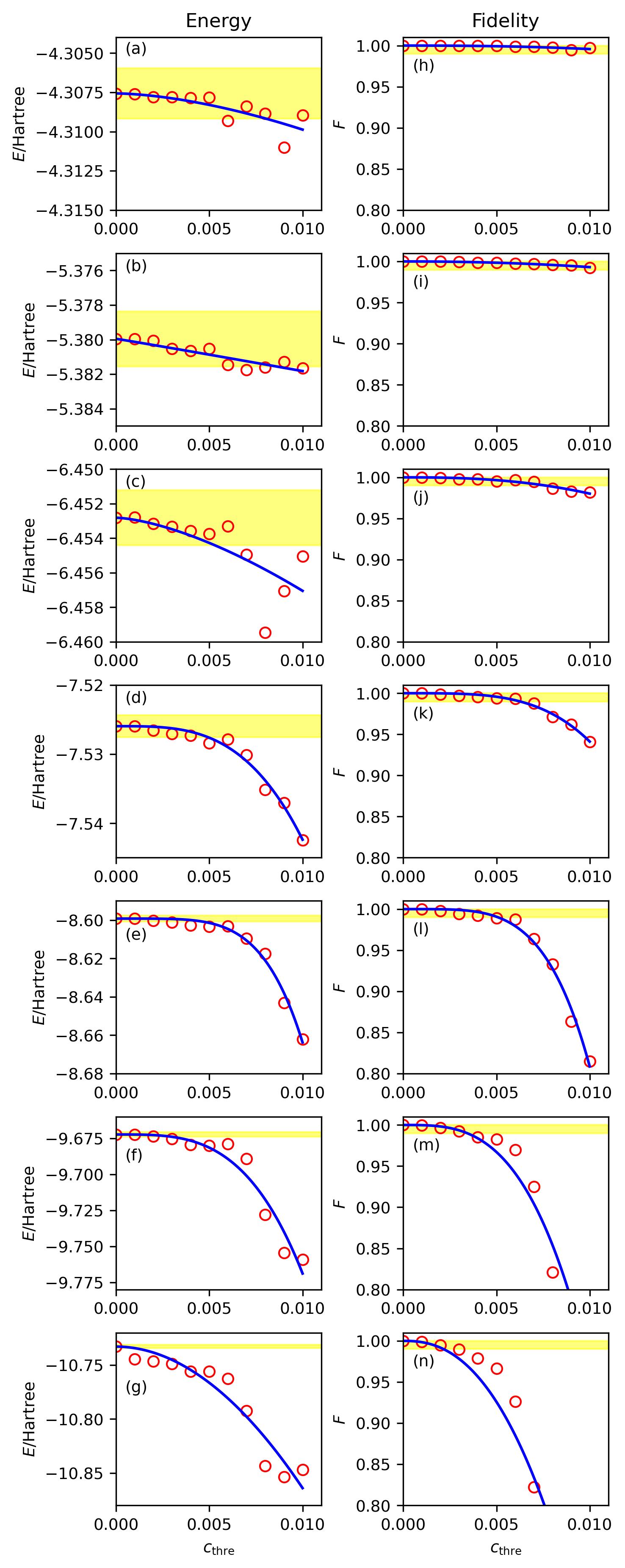}
    \caption{Dependence of the threshold value ($c_\text{thre}$) on the ground-state energy (a--g) and ground-state fidelity (h--n) for H$_8$ (a and h), H$_{10}$ (b and i), H$_{12}$ (c and j), H$_{14}$ (d and k), H$_{16}$ (e and l), H$_{18}$ (f and m), and H$_{20}$ (g and n) in the CMO-based wave function expansion. Red open circles represent calculated energies and fidelities, and blue curves specify the fitted curve obtained using eqs (\ref{eq6}) and (\ref{eq7}) for energies and fidelities, respectively. The yellow shaded area represents the region where the energy error $|\Delta E| \le 1.0\ \mathrm{kcal\ mol^{-1}}$ or the fidelity $F \ge 0.99$.}
    \label{fig:fig3}
\end{figure}

\begin{figure}
    \centering   
    \includegraphics[width=0.9\linewidth]{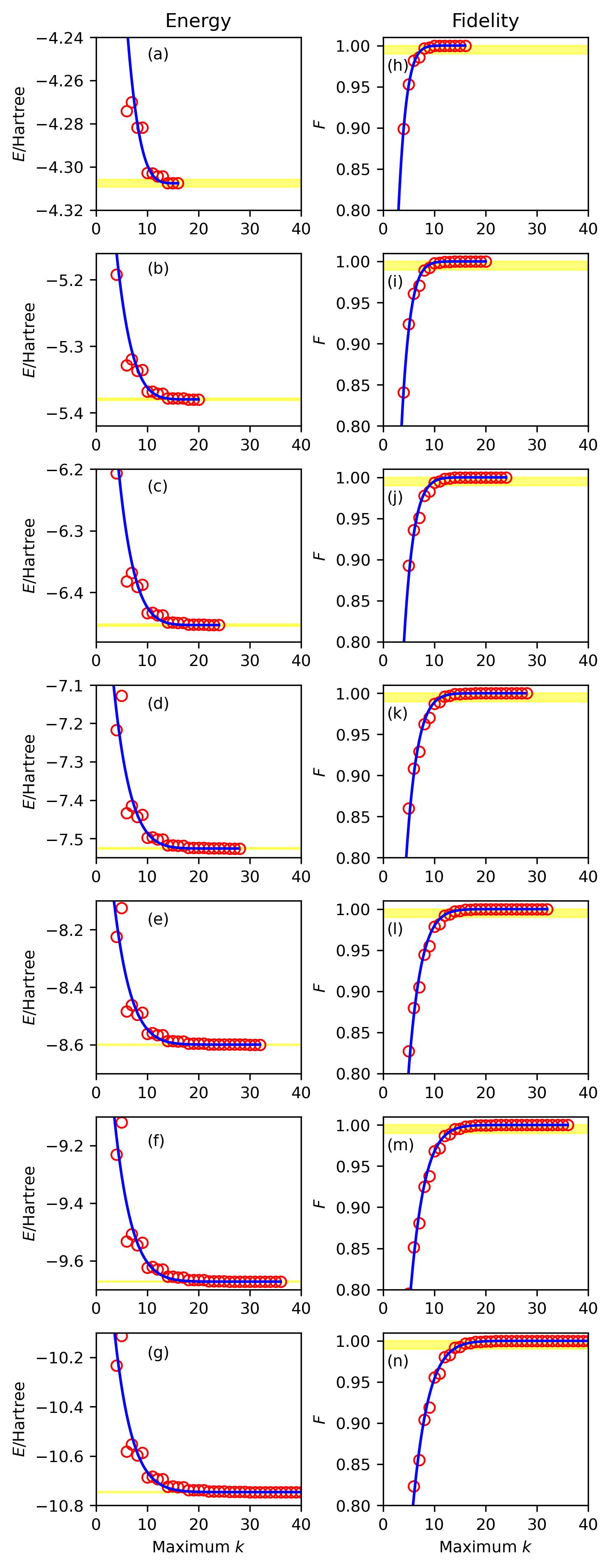}
    \caption{Dependence of the maximum locality value ($k_\text{max}$) on the ground-state energy (a--g) and ground-state fidelity (h--n) for H$_8$ (a and h), H$_{10}$ (b and i), H$_{12}$ (c and j), H$_{14}$ (d and k), H$_{16}$ (e and l), H$_{18}$ (f and m), and H$_{20}$ (g and n) in the LMO-based wave function expansion. Red open circles represent calculated energies and fidelities, and blue curves specify the fitted curve obtained using eqs (\ref{eq8}) and (\ref{eq9}) for energies and fidelities, respectively. The yellow shaded area represents the region where the energy error $|\Delta E| \le 1.0\ \mathrm{kcal\ mol^{-1}}$ or the fidelity $F \ge 0.99$.}
    \label{fig:fig4}
\end{figure}

Next, we study the effect of Hamiltonian truncation as a function of hydrogen chain length. We consider H$_8$, H$_{10}$, H$_{12}$, H$_{14}$, H$_{16}$, H$_{18}$, and H$_{20}$, to examine the dependence of the $E_\text{GS;truncated}$ and $F$, on the truncation threshold. We first assess the accuracy of the DMRG ground state by comparing the DMRG energies with the full-CI values for the H$_8$--H$_{14}$ systems while increasing the bond dimension from 100 to 1600. The results are summarized in Table S1 of the Supporting Information. With a bond dimension of 1600, the energy error is below 10$^{-9}$ Hartree for all cases except H$_{14}$ with CMO. Because CMOs are delocalized over the entire molecule, the required bond dimension generally increases with system size. In contrast, since LMOs are spatially localized, the accuracy of the ground state is maintained even with a relatively small bond dimension.

The effects of Hamiltonian truncation on the ground-state energies and fidelities for H$_8$--H$_{20}$ calculated using DMRG are summarized in Figures \ref{fig:fig3} and \ref{fig:fig4} for the CMO and LMO bases, respectively. The same calculations are also performed at the full-CI level for H$_8$--H$_{14}$, obtaining the differences between DMRG and full-CI in energy less than $2 \times 10^{-6}$ Hartree and in fidelity less than $4 \times 10^{-5}$.
To determine appropriate threshold values for Hamiltonian truncation that provide a sufficiently accurate description of the ground state in terms of energy (i.e., $|\Delta E| \le 1.0\ \mathrm{kcal\ mol^{-1}}$) and fidelity (i.e., $F \ge 0.99$), we perform curve-fitting analyses of the scatter plots of threshold value vs. $E_\text{GS;truncated}$ and $F$. The following fitting functions are used for the energy and fidelity data. 

For CMO with $c_\text{thre}$-based truncation:
\begin{eqnarray}
    E_\text{GS;truncated} = E_\text{GS;original} + a(c_\text{thre})^b
    \label{eq6}
\end{eqnarray}
\begin{eqnarray}
    F = 1.0 - a(c_\text{thre})^b 
    \label{eq7}
\end{eqnarray}

For LMO with $k_\text{max}$-based truncation: 
\begin{eqnarray}
    E_\text{GS;truncated} = E_\text{GS;original} + a(2n - k_\text{max})^b
    \label{eq8}
\end{eqnarray}
\begin{eqnarray}
    F = 1.0 - a(2n - k_\text{max})^b
    \label{eq9}
\end{eqnarray}

Here, $a$ and $b$ are fitting parameters, and $n$ is the number of hydrogen atoms. Note that these functions were chosen empirically, and there is no theoretical justification for assuming that $E_\text{GS;truncated}$ and $F$ follows these forms. Parameter fitting was performed using the \texttt{curve\_fit} function in SciPy \cite{SciPy}. The fitted parameters are summarized in Table \ref{tab:table1}, and fitted curves are also shown in Figures \ref{fig:fig3} and \ref{fig:fig4}. 

\begin{table}
\caption{\label{tab:table1} Fitted parameters and coefficient of determination for the threshold dependence of $E_\text{GS;truncated}$ and $F$, obtained using eqs (6) and (8), and eqs (7) and (9), respectively.}
\begin{tabular*}{\linewidth}{@{\extracolsep{\fill}}cccrr}
\hline
Fitting function & $n$ & $a$ & \multicolumn{1}{c}{$b$} & \multicolumn{1}{c}{$R^2$}\\
\hline
(6) &  8 & $-4.7036 \times 10^0$    & 1.6549 & 0.6100 \\
(6) & 10 & $-2.0934 \times 10^{-1}$ & 1.0246 & 0.8437 \\
(6) & 12 & $-4.9976 \times 10^0$    & 1.5362 & 0.5337 \\
(6) & 14 & $-6.4439 \times 10^4$    & 3.2968 & 0.9802 \\
(6) & 16 & $-3.2063 \times 10^8$    & 4.8481 & 0.9861 \\
(6) & 18 & $-5.9422 \times 10^5$    & 3.3945 & 0.9396 \\
(6) & 20 & $-1.1206 \times 10^3$    & 1.9660 & 0.9139 \\
(7) &  8 & $3.0217 \times 10^2$     & 2.4269 & 0.7722 \\
(7) & 10 & $1.5405 \times 10^2$     & 2.1746 & 0.9659 \\
(7) & 12 & $2.3620 \times 10^3$     & 2.5382 & 0.9427 \\
(7) & 14 & $2.1704 \times 10^6$     & 3.7837 & 0.9880 \\
(7) & 16 & $1.0344 \times 10^8$     & 4.3665 & 0.9886 \\
(7) & 18 & $6.3875 \times 10^5$     & 3.1673 & 0.9553 \\
(7) & 20 & $2.3238 \times 10^4$     & 2.3885 & 0.9307 \\
(8) &  8 & $4.2836 \times 10^{-6}$  & 4.2357 & 0.8187 \\
(8) & 10 & $8.0634 \times 10^{-8}$  & 5.3467 & 0.8345 \\
(8) & 12 & $1.1673 \times 10^{-9}$  & 6.4495 & 0.8493 \\
(8) & 14 & $1.3786 \times 10^{-11}$ & 7.5444 & 0.8620 \\
(8) & 16 & $1.3888 \times 10^{-13}$ & 8.6304 & 0.8725 \\
(8) & 18 & $1.2270 \times 10^{-15}$ & 9.7083 & 0.8812 \\
(8) & 20 & $9.6877 \times 10^{-18}$ & 10.7792 & 0.8883 \\
(9) &  8 & $6.2391 \times 10^{-11}$ & 8.5329 & 0.9961 \\
(9) & 10 & $2.0383 \times 10^{-13}$ & 9.8701 & 0.9939 \\
(9) & 12 & $3.8436 \times 10^{-15}$ & 10.5541 & 0.9919 \\
(9) & 14 & $2.5190 \times 10^{-16}$ & 10.8674 & 0.9913 \\
(9) & 16 & $2.9915 \times 10^{-17}$ & 11.0480 & 0.9916 \\
(9) & 18 & $4.7618 \times 10^{-18}$ & 11.1865 & 0.9925 \\
(9) & 20 & $8.5529 \times 10^{-19}$ & 11.3254 & 0.9934 \\
\hline
\end{tabular*}
\end{table}

The fitting results for the energy values involve relatively large errors ($R^2$ = 0.6100 and 0.5337 for H$_8$ and H$_{12}$, respectively, in the CMO basis), since the energy sometimes oscillate as a function of the threshold value. However, the ability to fit the energy with a function suggests that algorithmic error mitigation \cite{Endo-2019} can be used to estimate the eigenvalues of the original Hamiltonian. By contrast, the fidelity appears to be reproduced reasonably well by the fitting in both the CMO and LMO bases except for H$_8$ in the CMO basis ($R^2$ = 0.7722). From the intersections of the fitted curves with $F = 0.99$, the threshold values are estimated to be $c_\text{thre}$ = 0.014254 (H$_8$), 0.011865 (H$_{10}$), 0.007639 (H$_{12}$), 0.006262 (H$_{14}$), 0.005087 (H$_{16}$), 0.003433 (H$_{18}$), and 0.002161 (H$_{20}$), for the CMO-based wave function expansion, and $k_\text{max}$ = 6.8471 (H$_8$), 7.8900 (H$_{10}$), 8.9911 (H$_{12}$), 10.1622 (H$_{14}$), 11.3636 (H$_{16}$), 12.5736 (H$_{18}$), and 13.7733 (H$_{20}$), for the LMO-based ones. 

\begin{figure}
    \centering   
    \includegraphics[width=\linewidth]{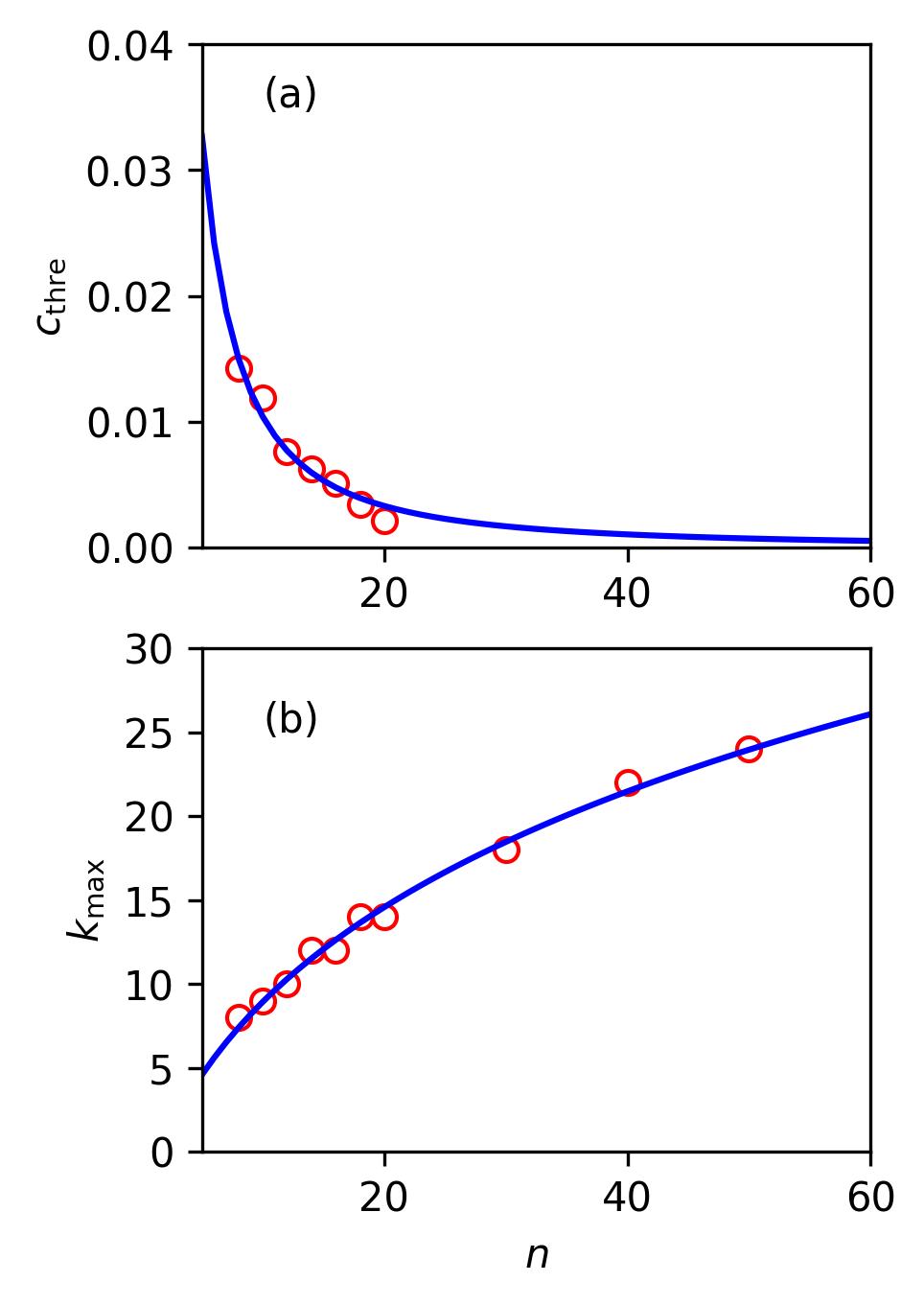}
    \caption{Dependence of the threshold value required for $F \ge 0.99$ on the number of hydrogen atoms $n$. Open red circles represent the threshold values obtained from the fitting functions in eqn (7) for the CMO basis, and the $k_\text{max}$ values achieving $F \ge 0.99$ in the LMO basis. The blue curves show the fitted functions, $c_\text{thre} = 1.3895 n^{-2.0707}$ for the CMO basis and $k_\text{max} = 1.8912 (\log n)^{1.8614}$. (a) Threshold value on the coefficient in the CMO-based wave function expansion. (b) Maximum locality value in the LMO-based wave function expansion.}
    \label{fig:fig5}
\end{figure}

To estimate the threshold for achieving $F \ge 0.99$ in larger systems in the CMO basis, we plotted the $c_\text{thre}$ values obtained from the fitting described above against the number of hydrogen atoms, excluded H$_8$, which had a small $R^2$ value, and then fitted the data using the function $c_\text{thre} = \alpha n^\beta$. The fitted parameters are $\alpha = 1.3895$ and $\beta = -2.0707$, with $R^2 = 0.9802$ (see Figure \ref{fig:fig5}a for details). To validate the fitting function, we performed a DMRG calculations for the H$_{30}$ molecule with $c_\text{thre}$ = 0.001214, obtaining $F$ = 0.992175. For the LMO basis, when the $k_\text{max}$ values obtained from the fitting were plotted against the number of hydrogen atoms, they appeared to be well approximated by a linear function. However, especially for long hydrogen chains, it seems unlikely that achieving $F \ge 0.99$ would require a linear increase in locality with system size. This is because, in the LMO basis, the operator locality $k$ is almost directly related to spatial distance, and correlations between electrons occupying molecular orbitals that are well separated in space are expected to be negligible. In fact, the linear function predicted that $k_\text{max}$ = 20, 26, and 32 would be required to achieve $F \ge 0.99$ for H$_{30}$, H$_{40}$, and H$_{50}$, respectively, but the DMRG calculations revealed that $F \ge 0.99$ is achieved at $k_\text{max}$ = 19, 22, and 24 for H$_{30}$, H$_{40}$, and H$_{50}$, respectively. We therefore plotted the values of $k$ required to actually achieve $F \ge 0.99$ against the number of hydrogen atoms and fitted them using the function $k = \alpha (\log n)^\beta$ for $n$ = 8, 10, 12, 14, 16, 18, 20, 30, 40, and 50, as shown in Figure \ref{fig:fig5}b. The fitted parameters are $\alpha = 1.8912$ and $\beta = 1.8614$, with $R^2 = 0.9995$. For the estimation of quantum gates required for Hamiltonian simulation, we adopt the following ceiling function $k_\text{max} = \lceil 1.8912 (\log n) ^{1.8614} \rceil$. We also confirm in the DMRG calculations that the estimated $k_\text{max}$ gives $F$ = 0.990479 for the H$_{60}$ system. 

\subsection{Quantum gate count analysis for Trotterized time evolution operator}

Lastly, we investigate how Hamiltonian truncation reduces the number of quantum gates required for the Trotter decomposition-based Hamiltonian simulation. Trotter decomposition \cite{Trotter-1959, Suzuki-1976} is a simple and well-studied approach for constructing a quantum circuits for real-time evolution. To keep the discussion as simple as possible, we evaluate the gate count for a single Trotter step ($M = 1$) in the first-order Trotter decomposition with $t = 1$, 
\begin{eqnarray}
    e^{-i\sum_j c_j P_j t} \approx \left[ \Pi_j e^{-i c_j P_jt/M} \right]^M.
    \label{eq11}
\end{eqnarray}
We construct the quantum circuit using a naive approach \cite{Whitfield-2011} and analyze the gate count without applying any quantum circuit optimization. Therefore, the quantum gate counts obtained in this study should be regarded as upper bounds. The numbers of one-qubit Clifford gates (H, $R_x(\pi/2)$, and $R_x(-\pi/2)$), CNOT gates, and $R_z(\theta)$ (non-Clifford) gates are counted separately. Two qubit gates such as the CNOT gate generally have higher error rates than one-qubit gates in quantum computation before FTQC, and the number of CNOT gates determines achievable computational scale. In the FTQC era, the number of non-Clifford gates, such as $R_z(\theta)$ gates, is the dominant contributor to the computational cost. 

\begin{figure*}
    \centering
    \includegraphics[width=\linewidth]{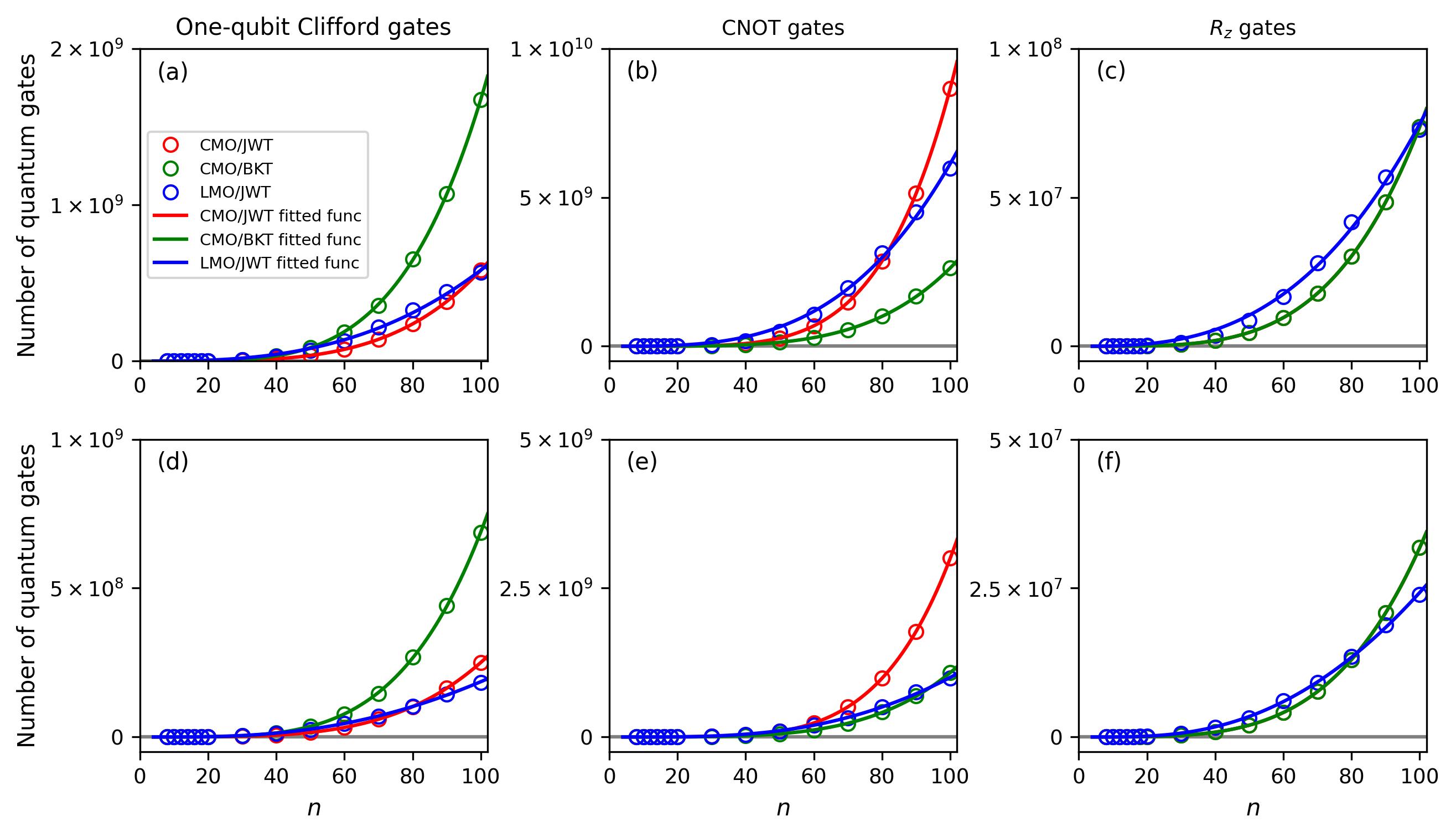}
    \caption{Number of quantum gates required for one Trotter step of Hamiltonian simulation with the first-order Trotter decomposition, using original (untruncated) Hamiltonian (a--c) and truncated Hamiltonian (d--f). One-qubit Clifford (H, $R_x(\pi/2)$, and $R_x(-\pi/2)$) gates (a and d), CNOT gates (b and e), and $R_z$ (non-Clifford) gates (c and f). The colored lines represent the results of fitting with a polylogarithmic function $\mathfrak{a} (\log n)^\mathfrak{b}$ for the LMO basis with Hamiltonian truncation, and a polynomial function $\mathfrak{a} n^\mathfrak{b}$ for others, where the fitted parameters are given in Tables II and III for original and truncated Hamiltonians, respectively. Note that the y-axis scales in subplots d--f are a half of those in subplots a--c.}
    \label{fig:fig6}
\end{figure*}

By assuming the scaling behaviors discussed in the previous section, the number of quantum gates required to achieve ground-state fidelity of $F \ge 0.99$, together with those for the original untruncated Hamiltonian, are plotted in Figure \ref{fig:fig6}, and the detailed gate counts and gate reduction ratio are given in Tables S2--S4 in the Supporting Information. The increase in the number of quantum gates with respect to the number of hydrogen atoms $n$ is well fitted by a polynomial function $\mathfrak{a} n^\mathfrak{b}$. For the LMO basis with operator locality-based Hamiltonian truncation, considering the fact that $k_\text{max}$ scales as a polylogarithmic function, we also fitted the data using $\mathfrak{a} (\log n)^\mathfrak{b}$, which gives slightly higher $R^2$ values. The fitted parameters are listed in Tables \ref{tab:table2} and \ref{tab:table3} for original and truncated Hamiltonians, respectively.
\begin{table}
\caption{\label{tab:table2} Fitted parameters and coefficient of determination for the quantum gate count of the original Hamiltonian.}
\begin{tabular*}{\linewidth}{@{\extracolsep{\fill}}ccrrr}
\hline
Quantum gates & MO/Encoding & \multicolumn{1}{c}{$\mathfrak{a}$} & \multicolumn{1}{c}{$\mathfrak{b}$} & \multicolumn{1}{c}{$R^2$}\\ 
\hline
\multirow{3}{*}{Clifford(1q)} & CMO/JW & 5.6187 & 4.0075 & 1.0000 \\ 
& CMO/BK & 4.7126 & 4.2749 & 0.9999 \\
& LMO/JW & 1236.2390 & 2.8363 & 0.9973 \\
\multirow{3}{*}{CNOT} & CMO/JW &1.0263 & 4.9632 & 1.0000 \\
& CMO/BK & 6.1744 & 4.3145 & 0.9999 \\
& LMO/JW & 2065.1637 & 3.2363 & 0.9973 \\
\multirow{3}{*}{$R_z$} & CMO/JW & 0.7682 & 3.9910 & 1.0000 \\
& CMO/BK & 0.7682 & 3.9910 & 1.0000 \\
& LMO/JW & 156.5605 & 2.8392 & 0.9974 \\
\hline
\end{tabular*}
\end{table}
\begin{table}
\caption{\label{tab:table3} Fitted parameters and coefficient for determination for the quantum gate count of the truncated Hamiltonian. Poly and polylog in the parenthesis indicate fitting with a polynomial function $\mathfrak{a}n^\mathfrak{b}$ and a polylogarithmic function $\mathfrak{a}(\log n)^\mathfrak{b}$, respectively.}
\begin{tabular*}{\linewidth}{@{\extracolsep{\fill}}ccrrr}
\hline
Quantum gates & MO/Encoding & \multicolumn{1}{c}{$\mathfrak{a}$} & \multicolumn{1}{c}{$\mathfrak{b}$} & \multicolumn{1}{c}{$R^2$} \\
\hline
\multirow{3}{*}{Clifford(1q)}
& CMO/JW          &   2.1930 &  4.0286 & 1.0000 \\
& CMO/BK          &   1.9040 &  4.2790 & 0.9999 \\
& LMO/JW(poly)    & 527.3646 &  2.7749 & 0.9986 \\
& LMO/JW(polylog) &   2.0098 & 12.0099 & 0.9995 \\
\multirow{3}{*}{CNOT}
& CMO/JW          &   0.3164 &  4.9885 & 1.0000 \\
& CMO/BK          &   2.3290 &  4.3332 & 0.9999 \\
& LMO/JW(poly)    & 473.6613 &  3.1636 & 0.9985 \\
& LMO/JW(polylog) &   0.6963 & 13.8067 & 0.9992 \\
\multirow{3}{*}{$R_z$} 
& CMO/JW          &  0.3060 &  4.0083 & 1.0000 \\
& CMO/BK          &  0.3060 &  4.0083 & 1.0000 \\
& LMO/JW(poly)    & 71.4830 &  2.7670 & 0.9987 \\
& LMO/JW(polylog) &  0.2775 & 11.9739 & 0.9995 \\
\hline
\end{tabular*}

\end{table}
With the Hamiltonian truncation, the number of quantum gate grows as $\mathcal{O}(n^4)$--$\mathcal{O}(n^5)$ for the CMO basis, and $\mathcal{O}((\log n)^{12})$--$\mathcal{O}((\log n)^{14})$ for the LMO basis. It is interesting that Hamiltonian term truncation does not alter the scaling behavior of the number of quantum gates but only reduces the prefactor in the CMO-based wave function expansion. However, this result is consistent with the observation that the reduction ratio of the quantum gate count defined as the number of quantum gates for the truncated Hamiltonian divided by that for the original Hamiltonian converges to a certain value around H$_{40}$. Note that the quantum gate counts for Hamiltonian simulation of the original Hamiltonian in the LMO basis scales approximately as $\mathcal{O}(n^3)$, because terms with coefficients smaller than $10^{-10}$ in in the second-quantized Hamiltonian are neglected.

From these results, we obtain the following important insights. (I) The number of $R_z(\theta)$ gates is equal to the number of terms in  the qubit Hamiltonian. For small $n$, the LMO-based wave function expansion requires approximately three times as may qubit Hamiltonian terms as the CMO-based expansion. However, this ratio decreases as $n$ increases, and for $n \approx 80$, the LMO-based wave function expansion requires less Hamiltonian terms than the CMO-based one. (II) As discussed in the previous sections, the number of original Hamiltonian terms in the LMO basis is about twice that in the CMO basis because of the difference in point-group symmetry. If the point-group symmetry were the same for both CMO and LMO, the difference between the two would be much smaller, and the crossing point will move to smaller $n$. (III) The number of one-qubit Clifford gates in the LMO basis with Hamiltonian truncation is comparable to that in the CMO basis. (IV) For small $n$, the number of CNOT gates in the CMO basis is smaller than that in the LMO basis, but the trend is reversed for $n \ge 50$. The number of CNOT gates required to simulate a single Pauli string is given by $2(k-1)$, where $k$ is the locality of the Pauli string. This indicates that, even after Hamiltonian truncation, many Hamiltonian terms with large $k$ remains in the CMO basis. In contrast, in the LMO-based wave function expansion, the qubit Hamiltonian contains at most polylog($n$)-local terms, which suppress the increase in the number of CNOT gates. 

The above findings are obtained by applying JW for both CMO and LMO. The number of CNOT gates scales $\mathcal{O}(N^5)$ in JW, where $N$ is the number of qubits. This is because Hamiltonian contains $\mathcal{O}(N^4)$ terms and the quantum circuit for real-time evolution for each Pauli string requires $\mathcal{O}(N)$ of CNOT gates. It is known that the scaling of the number of CNOT gates improves to $\mathcal{O}(N^4 \log N)$ by using Bravyi--Kitaev (BK) transformation  \cite{Seeley-2012, Tranter-2018} instead of JW. Figure \ref{fig:fig6} also includes fermion--qubit encoding method dependence on the quantum gate counts in CMO-based wave function expansion with $c_\text{thre}$-based Hamiltonian term truncation. As expected, the number of CNOT gates decreases drastically for large $n$ by applying BK in the CMO basis, but the resulting values are still comparable to those in the LMO basis. Interestingly, the number of one-qubit Clifford gates increases by adopting BK. This is because the number of Pauli-X and Y terms in Pauli string is always 2 (one-electron terms) or 4 (two-electron terms) in JW, whereas that in BK is not necessary to be 2 or 4. In fact, in BK, up to 24 Pauli-X and Y are involved in Pauli string in the H$_{100}$ system.   

Our resource estimation indicates that megaquop machines \cite{Preskill-2025} would be required to perform Trotter decomposition-based Hamiltonian simulation for hydrogen chains with $n \ge 18$ under a naive implementation. In the near term, early-FTQC technologies such as space-time efficient analog rotation (STAR) architectures \cite{Akahoshi-2024, Toshio-2025, Toshio-2026, ismail2026fast} may offer a plausible route to addressing this challenge. The STAR architecture is particularly well suited to one-qubit rotation gates with small rotational angles. To assess its potential impact, we analyze the distributions of rotational angles for the $R_z(\theta)$ gates in the H$_{50}$ system for both the CMO and LMO bases, with and without Hamiltonian truncation. The results are shown in Figure S1 in the Supporting Information. The distribution of rotation angles is very similar before and after Hamiltonian truncation, with the difference appearing in the number of $R_z(\theta)$ gates with very small rotational angles ($|\theta| \le 0.005\times \pi$ radian). Even after Hamiltonian truncation, more than 99\% of the $R_z(\theta)$ gates still satisfy $|\theta| \le 0.005 \times \pi$ radian, suggesting that the cost of Hamiltonian simulation could be reduced substantially further by adopting the STAR architecture. Notably, most of the $R_z(\theta)$ gates with small rotation angles originate from the two-electron operators describing electron--electron interactions. We do not expect the magnitude of the two-electron integrals to increase drastically even for more complex three-dimensional molecules or for systems containing heavy atoms, and therefore we believe that the conclusions of this study are broadly applicable. In fact, for the (113e, 76o) active space of FeMoco \cite{Li-2019}, the maximum absolute value of the two-electron integrals is 0.892606, which is of the same oeder of magnitudes as the corresponding values in H$_{50}$: 0.137778 for CMO and 0.658402 for LMO.

\section{Conclusion}
In this work, we investigated the computational resource required for Trotter decomposition-based Hamiltonian simulation of one-dimensional hydrogen chain systems, with a particular focus on the effect of molecular orbital localization and Hamiltonian truncation. We found that Hamiltonian truncation based on operator coefficiens is well suited to CMO-based wave function expansions, whereas truncation based on operator locality is more suitable for LMO-based expansions. Based on the fitting of the threshold dependence of the computed ground-state fidelity and the distance dependence of the intermolecular interaction energy obtained with the supramolecular approach, we estimated the threshold required to achieve ($F \ge 0.99$). We found that, in the CMO basis, $c_\text{thre}$ scales as $n^{-2.0707}$, where $n$ is the number of hydrogen atoms, whereas in the LMO basis, $k_\text{max}$ needs to increase polylogarithmically with the system size. This difference is also strongly reflected in the system-size dependence of the number of quantum gates required for Hamiltonian simulation: the gate count grows as $\mathcal{O}(n^4)$--$\mathcal{O}(n^5)$ in the CMO basis, but only $\mathcal{O}((\log n)^{12})$--$\mathcal{O}((\log n)^{14})$ in the LMO basis.

The present study focused on simple one-dimensional hydrogen chain systems, in which operator locality in the LMO basis directly correlated with the relative distance between two molecular orbitals. For more complex three-dimensional molecules, distance-based Hamiltonian truncation may be more appropriate than the locality-based one. A systematic study of computational resources for other systems with localized orbitals remains an important subject for future work. 

For the practical application of quantum computing to industrially relevant molecules, the preparation of appropriate quantum states on a quantum computer may represent another major bottleneck \cite{Lee-2023}, and resource estimation for state preparation is therefore essential for evaluating the end-to-end computational cost \cite{Fomichev-2024}. We hope that the insights obtained in this study will provide a useful foundation for the resource-aware design of quantum algorithms for chemically relevant systems. 

\section{Acknowledgments}
We thank Mekena McGrew and Unpil Baek for helpful discussions. This work was partially supported by the Center of Innovation for Sustainable Quantum AI (Grant Number JPMJPF2221) from Japan Science and Technology Agency (JST), Japan.

\section{Data availability}
The data that support the findings of this study are available from the corresponding author upon reasonable request.

\bibliography{ref}
\bibliographystyle{unsrt}

\clearpage
\onecolumngrid
\section{Supporting Information}

\renewcommand{\thepage}{S\arabic{page}}
\setcounter{page}{1}
\renewcommand{\thefigure}{S\arabic{figure}}
\setcounter{figure}{0}
\renewcommand{\thetable}{S\arabic{table}}
\setcounter{table}{0}

\subsection{Comparison of full-CI and DMRG energies}

\begin{table*}[h]
\caption{\label{table_s1} Full-CI energies and DMRG energy errors with different bond dimension values $D$.}
\begin{tabular*}{\linewidth}{@{\extracolsep{\fill}}cccccccc}
\hline
\multirow{2}{*}{System} & \multirow{2}{*}{MO} & $E_\text{full-CI}$ & \multicolumn{5}{c}{$\Delta E_\text{DMRG}/\textrm{Hartree}$} \\
 & & /Hartree & $D = 100$ & $D = 200$ & $D = 400$ & $D = 800$ & $D = 1600$ \\
 \hline
 H$_8$    & CMO & $-4.3075716020$ & 0.0000000003 & 0.0000000003 & 0.0000000004 & 0.0000000005 & 0.0000000005 \\
 H$_8$    & LMO & $-4.3075716020$ & 0.0000000003 & 0.0000000003 & 0.0000000005 & 0.0000000004 & 0.0000000002 \\
 H$_{10}$ & CMO & $-5.3799547461$ & 0.0000830947 & 0.0000007828 & 0.0000000002 & 0.0000000002 & 0.0000000003 \\
 H$_{10}$ & LMO & $-5.3799547461$ & 0.0000000253 & 0.0000000001 & 0.0000000004 & 0.0000000002 & 0.0000000004 \\
 H$_{12}$ & CMO & $-6.4528158554$ & 0.0012660637 & 0.0001831498 & 0.0000056683 & 0.0000000139 & 0.0000000001 \\
 H$_{12}$ & LMO & $-6.4528158554$ & 0.0000007300 & 0.0000000236 & 0.0000000002 & 0.0000000006 & 0.0000000004 \\
 H$_{14}$ & CMO & $-7.5259614683$ & 0.0055791372 & 0.0013853682 & 0.0002948113 & 0.0000239463 & 0.0000004552 \\
 H$_{14}$ & LMO & $-7.5259614683$ & 0.0000023308 & 0.0000000947 & 0.0000000020 & 0.0000000003 & 0.0000000004 \\
\hline
\end{tabular*}
\end{table*}

\subsection{Number of quantum gates required for Hamiltonian simulation}

\begin{table*}[h]
\caption{\label{table_s2} Number of quantum gates of the quantum circuit for the single Trotter step of Hamiltonian simulation with the first-order Trotter decomposition in the CMO basis with JW, with original (untruncated) and $c_\text{thre}$-based truncated Hamiltonians.}
\begin{tabular*}{\linewidth}{@{\extracolsep{\fill}}crrrrrrrccc}
\hline
\multirow{2}{*}{System} & \multicolumn{3}{c}{Original Hamiltonian} & \multicolumn{4}{c}{Truncated Hamiltonian} & \multicolumn{3}{c}{Reduction ratio\footnote{Calculated as the number of quantum gates for the truncated Hamiltonian divided by that for the original Hamiltonian.}} \\
 & \#Clifford(1q) & \#CNOT & \#$R_z$ & \multicolumn{1}{c}{$c_\text{thre}$} & \#Clifford(1q) & \#CNOT & \#$R_z$ & \#Clifford(1q) & \#CNOT & \#$R_z$ \\
\hline
H$_{8}$   &     19328 &      32160 &     2912 & 0.018741 &     12560 &      19552 &     1880 & 0.6498 & 0.6080 & 0.6456 \\
H$_{10}$  &     49440 &      96228 &     7150 & 0.011806 &     29648 &      54100 &     4278 & 0.5997 & 0.5622 & 0.5983 \\
H$_{12}$  &    105792 &     236064 &    14904 & 0.008094 &     61120 &     128512 &     8612 & 0.5777 & 0.5402 & 0.5778 \\
H$_{14}$  &    200480 &     504756 &    27734 & 0.005882 &    108352 &     252932 &    15062 & 0.5405 & 0.5011 & 0.5431 \\
H$_{16}$  &    347904 &     975808 &    47488 & 0.004461 &    180208 &     460464 &    24736 & 0.5180 & 0.4719 & 0.5209 \\
H$_{18}$  &    564640 &    1745764 &    76286 & 0.003495 &    283312 &     786036 &    38606 & 0.5018 & 0.4503 & 0.5061 \\
H$_{20}$  &    870080 &    2941056 &   116600 & 0.002810 &    423936 &    1266536 &    57268 & 0.4872 & 0.4306 & 0.4911 \\
H$_{30}$  &   4548704 &   21931868 &   595186 & 0.001214 &   2037600 &    8306212 &   268766 & 0.4480 & 0.3787 & 0.4516 \\
H$_{40}$  &  14599824 &   91451376 &  1888208 & 0.000669 &   6351728 &   32889256 &   827188 & 0.4351 & 0.3596 & 0.4381 \\
H$_{50}$  &  35953376 &  277048700 &  4617898 & 0.000421 &  15410000 &   97012572 &  1990898 & 0.4286 & 0.3502 & 0.4311 \\
H$_{60}$  &  74941952 &  685536352 &  9581616 & 0.000289 &  31961264 &  236470584 &  4106972 & 0.4265 & 0.3449 & 0.4286 \\
H$_{70}$  & 139264016 & 1474413708 & 17747722 & 0.000210 &  59449504 &  506948572 &  7608458 & 0.4269 & 0.3438 & 0.4287 \\
H$_{80}$  & 237980816 & 2861677816 & 30254700 & 0.000159 & 101759920 &  984428048 & 12984464 & 0.4276 & 0.3440 & 0.4292 \\
H$_{90}$  & 381494000 & 5135078500 & 48408422 & 0.000125 & 163632128 & 1772927580 & 20831138 & 0.4289 & 0.3453 & 0.4303 \\
H$_{100}$ & 581495952 & 8660502864 & 73676352 & 0.000100 & 250208256 & 3001665648 & 31794360 & 0.4303 & 0.3466 & 0.4315 \\
\hline
\end{tabular*}
\end{table*}

\begin{table*}[h]
\caption{\label{table_s3} Number of quantum gates of the quantum circuit for the single Trotter step of Hamiltonian simulation with the first-order Trotter decomposition in the CMO basis with BK, with original (untruncated) and  $c_\text{thre}$-based truncated Hamiltonians.}
\begin{tabular*}{\linewidth}{@{\extracolsep{\fill}}crrrrrrrccc}
\hline
\multirow{2}{*}{System} & \multicolumn{3}{c}{Original Hamiltonian} & \multicolumn{4}{c}{Truncated Hamiltonian} & \multicolumn{3}{c}{Reduction ratio\footnote{Calculated as the number of quantum gates for the truncated Hamiltonian divided by that for the original Hamiltonian.}} \\
 & \#Clifford(1q) & \#CNOT & \#$R_z$ & \multicolumn{1}{c}{$c_\text{thre}$} & \#Clifford(1q) & \#CNOT & \#$R_z$ & \#Clifford(1q) & \#CNOT & \#$R_z$  \\
\hline  
H$_{8}$   &      22368 &      30480 &     2912 & 0.018741 &     14040 &      19026 &     1880 & 0.6277 & 0.6242 & 0.6456 \\
H$_{10}$  &      73520 &      99060 &     7150 & 0.011806 &     42536 &      56904 &     4278 & 0.5786 & 0.5744 & 0.5983 \\
H$_{12}$  &     167472 &     227600 &    14904 & 0.008094 &     93468 &     126020 &     8612 & 0.5581 & 0.5537 & 0.5778 \\
H$_{14}$  &     327056 &     440560 &    27734 & 0.005882 &    171780 &     231288 &    15062 & 0.5252 & 0.5250 & 0.5431 \\
H$_{16}$  &     536768 &     751712 &    47488 & 0.004461 &    269200 &     375508 &    24376 & 0.5015 & 0.4995 & 0.5209 \\
H$_{18}$  &     985048 &    1392744 &    76286 & 0.003495 &    476984 &     674996 &    38606 & 0.4842 & 0.4847 & 0.5061 \\
H$_{20}$  &    1626640 &    2314248 &   116600 & 0.002810 &    759656 &    1077788 &    57268 & 0.4670 & 0.4657 & 0.4911 \\
H$_{30}$  &    9246352 &   13249060 &   595186 & 0.001214 &   3969276 &    5673794 &   268766 & 0.4293 & 0.4282 & 0.4516 \\
H$_{40}$  &   33503296 &   50064206 &  1888208 & 0.000669 &  13875120 &   20738022 &   827188 & 0.4141 & 0.4142 & 0.4381 \\
H$_{50}$  &   86813500 &  130705694 &  4617898 & 0.000421 &  35270772 &   53258372 &  1990898 & 0.4063 & 0.4075 & 0.4311 \\
H$_{60}$  &  185757740 &  280409490 &  9581616 & 0.000289 &  75890432 &  114159674 &  4106972 & 0.4085 & 0.4071 & 0.4286 \\
H$_{70}$  &  354767092 &  549741622 & 17747722 & 0.000210 & 145194680 &  225221104 &  7608458 & 0.4093 & 0.4097 & 0.4287 \\
H$_{80}$  &  654300264 & 1019081678 & 30254700 & 0.000159 & 267527244 &  416538758 & 12984464 & 0.4078 & 0.4087 & 0.4292 \\
H$_{90}$  & 1069340408 & 1673896642 & 48408422 & 0.000125 & 441452820 &  686575456 & 20831138 & 0.4128 & 0.4102 & 0.4303 \\
H$_{100}$ & 1672459956 & 2622899438 & 73676352 & 0.000100 & 686451956 & 1078624312 & 31794360 & 0.4104 & 0.4112 & 0.4315 \\
\hline
\end{tabular*}
\end{table*}

\begin{table*}[h]
\caption{\label{table_s4} Number of quantum gates of the quantum circuit for the single Trotter step of Hamiltonian simulation with the first-order Trotter decomposition in the LMO basis with JW, with original (untruncated) and $k_\text{max}$-based truncated Hamiltonians.}
\begin{tabular*}{\linewidth}{@{\extracolsep{\fill}}crrrrrrrccc}
\hline
\multirow{2}{*}{System} & \multicolumn{3}{c}{Original Hamiltonian} & \multicolumn{4}{c}{Truncated Hamiltonian} & \multicolumn{3}{c}{Reduction ratio\footnote{Calculated as the number of quantum gates for the truncated Hamiltonian divided by that for the original Hamiltonian.}} \\
 & \#Clifford(1q) & \#CNOT & \#$R_z$ & $k_\text{max}$ & \#Clifford(1q) & \#CNOT & \#$R_z$ & \#Clifford(1q) & \#CNOT & \#$R_z$ \\
\hline
H$_{8}$   &     38528 &      64192 &     5792 &  8 &     26848 &     37856 &     4304 & 0.6068 & 0.5897 & 0.7431 \\
H$_{10}$  &     98352 &     191700 &    14210 &  9 &     63152 &     98556 &     9702 & 0.6421 & 0.5141 & 0.6828 \\
H$_{12}$  &    211200 &     471792 &    29736 & 10 &    127412 &    217856 &    19000 & 0.6036 & 0.4618 & 0.6390 \\
H$_{14}$  &    399888 &    1007532 &    55306 & 12 &    259712 &    510060 &    37466 & 0.6495 & 0.5062 & 0.6774 \\
H$_{16}$  &    694720 &    1949648 &    94808 & 12 &    385120 &    773064 &    55188 & 0.5544 & 0.3965 & 0.5821 \\
H$_{18}$  &   1127312 &    3486540 &   152290 & 14 &    673840 &   1529380 &    94566 & 0.5977 & 0.4387 & 0.6210 \\
H$_{20}$  &   1737792 &    5874656 &   232864 & 14 &    910800 &   2100656 &   127320 & 0.5241 & 0.3576 & 0.5468 \\
H$_{30}$  &   9047632 &   43444476 &  1184114 & 18 &   4046672 &  11863852 &   548282 & 0.4473 & 0.2731 & 0.4630 \\
H$_{40}$  &  28573728 &  175540560 &  3698164 & 22 &  11832688 &  42039856 &  1575584 & 0.4141 & 0.2395 & 0.4260 \\
H$_{50}$  &  67212944 &  488951612 &  8649138 & 24 &  24118624 &  94142116 &  3187638 & 0.3588 & 0.1925 & 0.3685 \\
H$_{60}$  & 130006720 & 1064596176 & 16679048 & 27 &  45968528 & 201582592 &  6033040 & 0.3536 & 0.1894 & 0.3617 \\
H$_{70}$  & 218048544 & 1956279268 & 27935846 & 28 &  69504672 & 317887740 &  9106418 & 0.3188 & 0.1625 & 0.3260 \\
H$_{80}$  & 325353616 & 3140481784 & 41681792 & 30 & 102944928 & 504334992 & 13464288 & 0.3164 & 0.1606 & 0.3230 \\
H$_{90}$  & 443553584 & 4514072396 & 56878210 & 32 & 144042288 & 751505188 & 18821786 & 0.3247 & 0.1665 & 0.3309 \\
H$_{100}$ & 566709104 & 5983698720 & 72786872 & 33 & 183195712 & 985578048 & 23956016 & 0.3233 & 0.1647 & 0.3291 \\
\hline
\end{tabular*}
\end{table*}

\begin{figure}
    \centering
    \includegraphics[width=\linewidth]{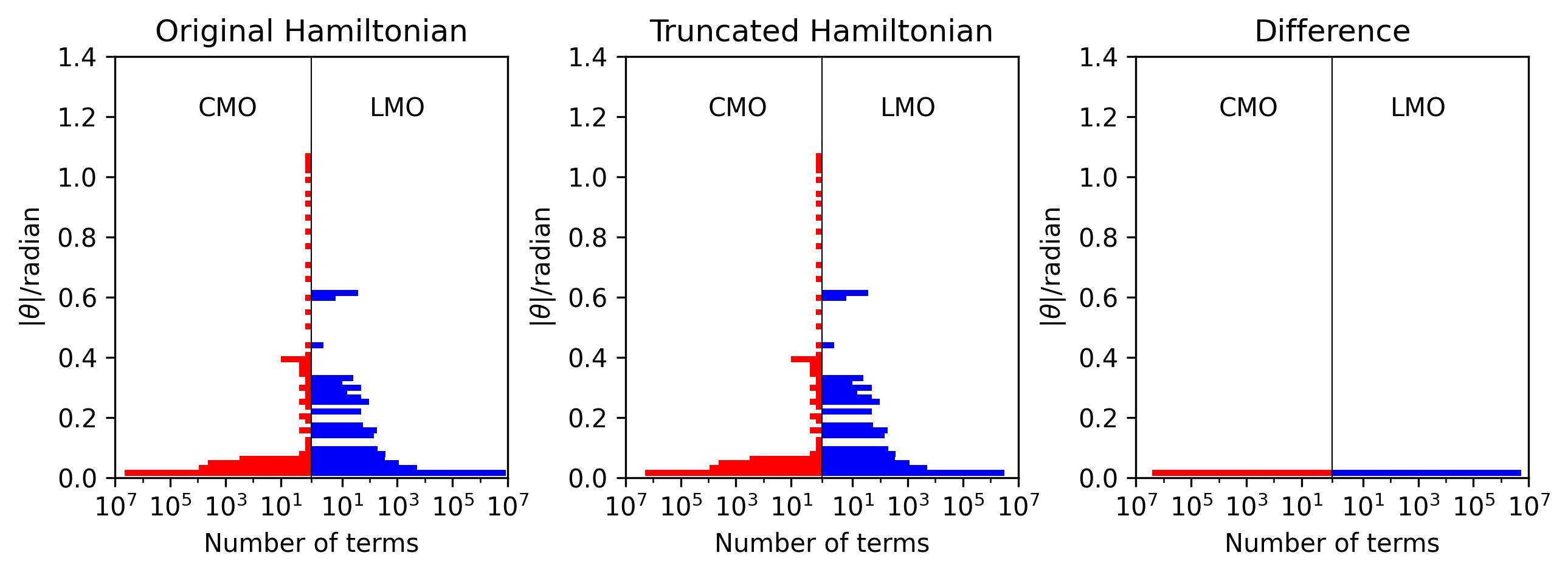}
    \caption{Distribution of the rotational angles of the $R_z(\theta)$ gates in the quantum circuit for a single Trotter step of Hamiltonian simulation of the H$_{50}$ molecule using first-order Trotter decomposition with $t = 1$ , based on original (left) and truncated (center) Hamiltonians. The right panel shows the difference in the number of terms between the original and truncated Hamiltonians.}
    \label{fig:figS1}
\end{figure}

\end{document}